\documentclass[twoside,fleqn]{article}
\usepackage{espcrc2}

\title{
How to compute one-loop Feynman 
diagrams in lattice QCD \\
with Wilson fermions 
}

\author{
  Giuseppe Burgio\address{Dipartimento di Fisica and INFN,
      Universit\`a degli Studi di Parma,
      Parma 43100, ITALIA},
  Sergio Caracciolo\address{Dipartimento di Fisica and INFN,
      Universit\`a degli Studi di Lecce,
      Lecce 73100, ITALIA}${}^,$\thanks{Speaker at the conference.}
 $\!$and
  Andrea Pelissetto\address{Dipartimento di Fisica and INFN,
      Universit\`a degli Studi di Pisa,
      Pisa 56100, ITALIA}
}

\begin{document}

\begin{abstract}
We describe an algebraic algorithm which allows to express every one-loop
lattice integral with gluon or Wilson-fermion propagators in terms of 
a small number of basic constants which can be computed with 
arbitrary high precision. Although the presentation is restricted to
four dimensions the technique can be generalized to 
every space dimension. 
We also give a method to express the lattice
free propagator for Wilson fermions in coordinate space as a linear function
of its values in eight points near the origin. This is an essential step
in order to apply the recent methods of L\" uscher and Weisz to 
higher-loop integrals with fermions.
\end{abstract}

\maketitle

\newcommand{\reff}[1]{(\ref{#1})}
\def\smfrac#1#2{{\textstyle\frac{#1}{#2}}}

\newcommand{\be}{\begin{equation}}
\newcommand{\ee}{\end{equation}}
\newcommand{\<}{\langle}
\renewcommand{\>}{\rangle}

\newcommand\FF{{\cal F}_\delta}

In \cite{CMP} we presented a general 
technique which allows to express every one-loop bosonic integral 
at zero external momentum 
in terms of two unknown basic quantities which could be computed
numerically with high precision. This method allows the complete 
evaluation of every diagram with gluon propagators. 

We have recently generalized \cite{BCP} this technique to deal with integrals
with both gluonic and fermionic propagators. For the fermions we use the 
Wilson action \cite{Wilson}. 
Notice that our method depends only on
the structure of the propagator and thus it can be applied in calculations
with the standard Wilson action as well as with the improved
clover action \cite{clover-action}. 
We show that every integral at zero external momentum 
can be expressed in terms of 
a small number of basic quantities (nine for purely fermionic integrals,
fifteen for integrals with bosonic and fermionic propagators). The 
advantage of this procedure is twofold: first of all every Feynman
diagram can be computed in a completely symbolic way making it easier
to perform checks and verify cancellations; moreover the basic 
constants can be easily computed with high precision and thus the 
numerical error on the final result can be reduced at will.
Although the presentation is restricted to
four dimensions the technique can be generalized to
every space dimension.

Define
\begin{eqnarray}
\lefteqn{\FF (p,q;n_x,n_y,n_z,n_t) =\, }\nonumber \\
&& \int_{-\pi}^\pi {d^4 k \over (2\pi)^4}
 {\hat{k}_x^{2 n_x} \hat{k}_y^{2 n_y} \hat{k}_z^{2 n_z} \hat{k}_t^{2 n_t}\over
     D_F^{p+\delta}(k,m_f) D_B^q(k,m_b)}
\end{eqnarray}
where $n_i$ are positive integers, $p$ and $q$ are arbitrary integers 
(not necessarily positive), $\hat{k}_\mu = 2 \sin(k_\mu/2)$ 
\begin{eqnarray}
D_B(k,m_b) &=& \hat{k}^2 + m_b^2 = \sum_i \hat{k}_i^2 + m_b^2
\; ;\label{bosonicprop}\\
D_F(k,m_f) &=& \sum_i \sin^2 k_i + {r^2_W\over 4} (\hat{k}^2)^2 + m_f^2
\label{Fermiprop}
\end{eqnarray}
is the denominator appearing in the 
propagator for Wilson fermions\footnote{To be precise, 
$D_F(k,m_f)$ is the denominator
in the propagator for Wilson fermions only for $m_f=0$.
For $m_f \not =0 $ the correct denominator would be
\be
\hat{D}_F(k,m_f) = \sum_i \sin^2 k_i + \left({r_W\over 2} \hat{k}^2 + m_f
    \right)^2 \;\; .
\label{trueFermiprop}
\ee
However in our discussion $m_f$ will only play the role of an infrared
regulator and thus it does not need to be the {\em true} fermion mass.
The definition \reff{Fermiprop} is easier to handle than
\reff{trueFermiprop}. }.
In the following when one of the arguments $n_i$ is zero it will be
omitted as an argument of $\cal F$.
The parameter $\delta$ 
is used in the intermediate steps of the calculation and will be set to zero
at the end.

To simplify the discussion we have only considered the case $r_W=1$
but the technique can be applied to every value of $r_W$. Moreover
we have restricted our attention to the massless case, i.e. we have 
considered the integrals $\FF$ in the limit $m_b=m_f\equiv m\to 0$.

We have developed a procedure that allows to compute iteratively
a generic ${\cal F}$, which is $\FF$ in the limit $\delta\to 0$, in
terms of a finite number of them: precisely  every ${\cal
F}(p,q;n_x,n_y,n_z,n_t)$ with $p>0$ and $q\le 0$  (purely fermionic
integrals) can be expressed in terms of ${\cal F}(1,0)$,
${\cal F}(1,-1)$,
${\cal F}(1,-2)$, ${\cal F}(2,0)$,
${\cal F}(2,-1)$, ${\cal F}(2,-2)$, ${\cal F}(3,-2)$, ${\cal
F}(3,-3)$ and
${\cal F}(3,-4)$; the integral ${\cal F}(2,0)$ appears only in
infrared-divergent integrals and we write it as
\be
   {\cal F}(2,0) =\, - {1\over 16 \pi^2}
   \left( \log m^2 + \gamma_E - F_0\right) + Y_0
\ee
where $Y_0$ is a numerical constant.
If $q>0$ the result contains three additional constants 
which we have chosen to be
\begin{eqnarray}
Y_1 &=& {1\over 8}\, {\cal F} (1,1;1,1,1) \nonumber \\
Y_2 &=& {1\over 16}\, {\cal F} (1,1;1,1,1,1) \nonumber \\
Y_3 &=& {1\over 16}\, {\cal F} (1,2;1,1,1) 
\end{eqnarray}
together with 
the bosonic quantities $Z_0$, $Z_1$ and $F_0-\gamma_E$.

As a first step in our procedure 
we express each integral $\FF(p,q;n_x,n_y,n_z,n_t)$ in terms
of integrals of the form $\FF(r,s)$ only. 
This is achieved by the use of the identity
\be
\sum_{i=1}^4 \hat{k}_i^2 = D_B(k,m) - m^2 
\ee
in the case one of the $n_i$'s is 1.
In the case one of the $n_i$'s is 2  we use instead
\begin{eqnarray}
\sum_{i=1}^4 \hat{k}_i^4 & = & 4\left[ D_B(k,m) - D_F(k,m)\right]\nonumber \\
 & &  +  \left[ D_B(k,m) - m^2\right]^2 
\end{eqnarray}
In all other cases we use an integration by parts. It is here that
we found convenient to have a non-vanishing $\delta$, otherwise
integration by parts in the case in which there is only one
fermionic propagator would produce a $\log D_F$. 

As a result we have
that
\begin{eqnarray}
\lefteqn{\FF (p,q;n_x,n_y,n_z,n_t) \,=\, } \nonumber \\
& &  \sum_{r=p-k+1}^p \sum_{s=q-k}^{q+k} a_{rs}(m,\delta) \FF(r,s)
\label{primariduzione}
\end{eqnarray}
where $k= (n_x + n_y + n_z + n_t)$, and $a_{rs}(m,\delta)$ is a polynomial
in $m^2$. 


At the  second step we express every $\FF(p,q)$ in terms of 
a finite number of them. We start by inserting the
trivial identity
\be
0 = 1 - {\sum \hat{k}_i^2 + m^2\over D_B(k,m)}
\ee
into the integral $\FF(p,q;1,1,1,1)$ to get 
\begin{eqnarray}
\lefteqn{0 = {\cal I}_1 (p,q) \equiv    }\nonumber\\ &&          
  \FF (p,q;1,1,1,1)- 4 \FF (p,q+1;2,1,1,1) \nonumber\\ &&          
  - m^2 \FF (p,q+1;1,1,1,1)    
\label{identity1}
\end{eqnarray}
Using the previous recursions we obtain from \reff{identity1} a non
trivial relation among the $\FF(r,s)$.  Starting from the analogous
relation obtained from the fermionic propagator we get a new set of
identities ${\cal I}_2 (p,q)$ which can be used to provide additional
relations between  the remaining integrals. We end up with the
result we have quoted above.

In order to numerically evaluate the eight basic purely-fermionic
integrals we considered the integrals
\be 
J_q = {\cal F}(1,-q)\qquad 6\leq q\leq 13 
\ee
which can be analytically expressed in terms of the elements in the fermionic 
basis. We computed 
\be
    J_{q,L} = {1\over L^4} \sum_k {D_B(k,0)^q\over D_F(k,0)}
\label{estrapolazionenumerica}
\ee
where $k$ runs over the points $k = (2\pi/L) (n_1+\smfrac{1}{2},
n_2+\smfrac{1}{2},n_3+\smfrac{1}{2},n_4+\smfrac{1}{2})$,
$0\le n_i < L$ for various values of $L$ between 50 and 100 and
extrapolated the results to infinite volume by using
\be
J_{q,L} \approx J_q \left( 1 + {A\over L^{2q+2}} \right)
\ee
The inversion of this set of equations provides a very accurate estimate of
the elements in the basis (see Table 1).
Analogously to compute $Y_0$, $Y_1$, $Y_2$ and $Y_3$,  we computed
numerically
${\cal F}(1,1,8)$, ${\cal F}(2,1,9)$, ${\cal F}(3,1,10)$ and 
${\cal F}(5,2,11)$ and then solved the corresponding equations.

A second important application of our method is connected with the
use  of coordinate-space methods for the evaluation of higher-loop 
Feynman diagrams. This technique, introduced by 
L\"uscher and Weisz \cite{LW}, is extremely powerful and allows  a
very precise determination of two- and higher-loop integrals.
One of the basic ingredients of this method is the computation of the 
free propagator in coordinate space. 
But this can be achieved by the use of our relations. Consider
\begin{eqnarray}
G(p,q;x) &=& \int {dk\over (2\pi)^4} 
{e^{ikx}\over D_F^p(k,m) D_B^q(k,m)} \\
&=& \int {dk\over (2\pi)^4} 
{\prod_\mu \cos k_\mu x_\mu \over D_F^p(k,m) D_B^q(k,m)} 
\label{GF}
\end{eqnarray}
Then express 
$\cos (k_\mu x_\mu)$ as a polynomial in $\hat{k}^2_\mu$. It follows that
$G$ is a linear combination of the $\cal F$ and therefore they can be expressed
on our basis.

Here, like in the bosonic case, the expression of $G$ in terms of the
basis become numerically unstable for $|x|\to\infty$: numerical
errors in the basis become amplified! This problem has a 
standard way  out: if the expressions are unstable going outward from 
the origin, they will be stable in the opposite direction: thus,
if we want to compute the propagator for $|x|<d$, for some fixed $d$, we 
choose eight points with $|x|\approx d$ (say $y_1,\ldots,y_8$) and 
then we express the propagator for $|x|<d$ in terms of 
$G(p,q;y_i)$, $i=1,\ldots,8$. The new expressions are numerically
stable: the numerical error on $G(p,q;y_i)$ gets 
{\em reduced} when we compute the propagator
for $|x|\to 0$. As noticed in \cite{LW} the instability of the recursion
can also be used to provide precise estimates for the basic constants.
We have thus used this method to obtain an independent numerical estimate 
of the eight purely fermionic and infrared finite constants in our
basis, considering the set of eight points 
$X^{(n)}\equiv \{(n,[0-3],0,0),(n+1,[0-3],0,0)\}$.

Because of better convergence, by using larger
negative values of $q$, one can obtain more precise estimates of 
$G(1,q;x)$ at the set of points $X^{(n)}$. 
Evaluating the integrals by computing discrete sums on finite
lattices, as before, with sizes $L=50$ -- $100$, we obtain  the
values of $G(1,-3;x)$ at the points
$X^{(26)}$ and from these numbers we obtained an independent
determination of the finite eight  fermionic constants (see Table
\ref{costanti_fermioniche}).

\begin{table}
\small
\begin{center}
\begin{tabular}{|c|l|}
\hline
${\cal F}(1,0)$ &0.08539036359532067913516702879 
\\
& 0.08539036359532067913516702676 
\\ \hline
${\cal F}(1,-1)$ &0.46936331002699614475347539703 
\\
& 0.46936331002699614475347539758 
\\ \hline
${\cal F}(1,-2)$ &3.39456907367713000586008689687 
\\
& 3.39456907367713000586008689071 
\\ \hline
${\cal F}(2,-1)$ &0.05188019503901136636490228763 
\\
& 0.05188019503901136636490228706 
 \\ \hline
${\cal F}(2,-2)$ &0.23874773756341478520233613924 
\\
& 0.23874773756341478520233613767 
 \\ \hline
${\cal F}(3,-2)$ &0.03447644143803223145396188143 
\\
& 0.03447644143803223145396188128 
 \\ \hline
${\cal F}(3,-3)$ &0.13202727122781293085314731096 
\\
& 0.13202727122781293085314731035 
 \\ \hline
${\cal F}(3,-4)$ &0.75167199030295682253543148585 
\\
& 0.75167199030295682253543148379 
\\ 
\hline
\end{tabular}
\end{center}
\caption{Numerical values of the constants appearing in the
fermionic  integrals evaluated by the two different methods 
described. The first line refers to the method which goes through
the evaluation of the integrals $J_q$, the second one goes through
the propagators in coordinate space. In all cases there  is an
agreement of at least 27 digits.  }
\label{costanti_fermioniche}
\end{table}

%
%
%

\end{document}